\documentclass[11pt,preprint]{aastex}
\newcommand{\co}{\rm CO}
\newcommand{\hh}{{\rm H}_2}

\newcommand{\kps}{\,\textstyle\rm{km~s}^{-1}}

\newcommand{\codm}{{\rm CO(4-3)}}
\newcommand{\cocm}{{\rm CO(3-2)}}
\newcommand{\cobm}{{\rm CO(2-1)}}
\newcommand{\coam}{{\rm CO(1-0)}}

\newcommand{\otz}{{\rm 1-0}}
\newcommand{\coc}{CO(3-2)\,}

\newcommand{\coa}{CO(1-0)\,}

\newcommand{\msun}{\,M_{\odot}}
\newcommand{\lsun}{\,L_{\odot}}

\newcommand{\jy}{\,\textstyle\rm{Jy}}

\newcommand{\Kkpspc}{\,\rm{K}\,\rm{km~s}^{-1}\,{\rm pc}^{2}}

\slugcomment{Accepted ApJL}

\shorttitle{CO(1-0) Observations of H-ATLAS SMGs}
\shortauthors{Frayer et al.}

\begin{document}

\title{GBT Zpectrometer CO(1-0) Observations of the Strongly-Lensed
 Submillimeter Galaxies from the {\em Herschel} ATLAS}

\author{D. T. Frayer\altaffilmark{1}, A. I. Harris\altaffilmark{2},
  A. J. Baker\altaffilmark{3}, R. J. Ivison\altaffilmark{4,5}, Ian
  Smail\altaffilmark{6}, M. Negrello\altaffilmark{7},
  R. Maddalena\altaffilmark{1}, I. Aretxaga\altaffilmark{8},
  M. Baes\altaffilmark{9}, M. Birkinshaw\altaffilmark{10},
  D. G. Bonfield\altaffilmark{11}, D. Burgarella\altaffilmark{12},
  S. Buttiglione\altaffilmark{13}, A. Cava\altaffilmark{14},
  D. L. Clements\altaffilmark{15}, A. Cooray\altaffilmark{16},
  H. Dannerbauer\altaffilmark{17}, A. Dariush\altaffilmark{18}, G. De
  Zotti\altaffilmark{19}, J. S. Dunlop\altaffilmark{20},
  L. Dunne\altaffilmark{21}, S. Dye\altaffilmark{18},
  S. Eales\altaffilmark{18}, J. Fritz\altaffilmark{9},
  J. Gonzalez-Nuevo\altaffilmark{22}, D. Herranz\altaffilmark{23},
  R. Hopwood\altaffilmark{7}, D. H. Hughes\altaffilmark{8},
  E. Ibar\altaffilmark{4}, M. J. Jarvis\altaffilmark{11},
  G. Lagache\altaffilmark{24}, L. L. Leeuw\altaffilmark{25,26},
  M. Lopez-Caniego\altaffilmark{23}, S. Maddox\altaffilmark{21},
  M. J. Micha{\l}owski\altaffilmark{20}, A. Omont\altaffilmark{27},
  M. Pohlen\altaffilmark{18}, E. Rigby\altaffilmark{21},
  G. Rodighiero\altaffilmark{28}, D. Scott\altaffilmark{29},
  S. Serjeant\altaffilmark{7}, D. J. B. Smith\altaffilmark{21},
  A. M. Swinbank\altaffilmark{6}, P. Temi\altaffilmark{30},
  M. A. Thompson\altaffilmark{11}, I. Valtchanov\altaffilmark{31},
  P. P. van der Werf\altaffilmark{32,5}, A. Verma\altaffilmark{33}}

\altaffiltext{1}{National Radio Astronomy Observatory, PO Box 2, Green
  Bank, WV 24944, USA}

\altaffiltext{2}{Department of Astronomy, University of Maryland,
College Park, MD 20742, USA}

\altaffiltext{3}{Department of Physics and Astronomy, Rutgers, the State
University of New Jersey, 136 Frelinghuysen Road, Piscataway, NJ
08854-8019, USA}

\altaffiltext{4}{UK Astronomy Technology Centre, Royal Observatory,
  Blackford Hill, Edinburgh EH9 3HJ, UK}

\altaffiltext{5}{Institute for Astronomy, University of Edinburgh,
Royal Observatory, Blackford Hill, Edinburgh EH9 3HJ, UK}

\altaffiltext{6}{Institute for Computational Cosmology, Physics
  Department, Durham University, South Road, Durham DH1 3LE, UK}

\altaffiltext{7}{Department of Physics and Astronomy, The Open
  University, Milton Keynes, MK7 6AA, UK}

\altaffiltext{8}{Instituto Nacional de Astrof\'{i}sica, \'{O}ptica y
Electr\'{o}nica Luis Enrique Erro, 1 Tonantzintla, Puebla, 72840,
Mexico}

\altaffiltext{9}{Sterrenkundig Observatorium, Universiteit Gent,
  Krijgslaan 281 S9, B-9000 Gent, Belgium}

\altaffiltext{10}{Department of Physics, University of Bristol,
  Tyndall Avenue, Bristol BS8 1TL, UK}

\altaffiltext{11}{Centre for Astrophysics Research, Science \&
Technology Research Institute, University of Hertfordshire, Hatfield,
Herts, AL10 9AB, UK}

\altaffiltext{12}{Laboratoire d'Astrophysique de Marseille, UMR6110
CNRS, 38 rue F. Joliot-Curie, 13388 Marseille, France}

\altaffiltext{13}{INAF-Osservatorio Astronomico di Padova, Vicolo
Osservatorio I-35122 Padova, Italy}

\altaffiltext{14}{Instituto de Astrof\'{i}sica de Canarias, C/V\'{i}a
L\'{a}ctea s Laguna, Spain}

\altaffiltext{15}{Physics Department, Imperial College London, Prince
Consort Road, London, SW7 2AZ, UK}

\altaffiltext{16}{Center for Cosmology, University of California,
Irvine, CA 92697, USA}

\altaffiltext{17}{AIM, CEA/DSM-CNRS-Universit Paris Diderot,
  DAPNIA/Service d'Astrophysique, CEA Saclay, Orme De Merisiers, 91191
  Gif-sur-Yvette, Cedex, France}

\altaffiltext{18}{School of Physics and Astronomy, Cardiff University,
Cardiff CF24 3AA, UK}

\altaffiltext{19}{INAF-Osservatorio Astronomico di Padova, Vicolo
Osservatorio 5, I-35122 Padova, Italy, and SISSA, Via Bonomea 265,
I-34136 Trieste, Italy}

\altaffiltext{20}{Scottish Universities Physics Alliance, Institute
  for Astronomy, University of Edinburgh, Royal Observatory, Edinburgh,
  EH9 3HJ, UK}

\altaffiltext{21}{School of Physics and Astronomy, University of
Nottingham, Nottingham, NG7 2RD, UK}

\altaffiltext{22}{Scuola Internazionale Superiore di Studi Avanzati,
via Beirut 2-4, 34151 Triest, Italy}

\altaffiltext{23} {Instituto de Fisica de Cantabria (CSIC-UC),
  Avda. los Castros s/n, 39005 Santander, Spain}

\altaffiltext{24}{Institut d'Astrophysique Spatiale, Universit
Paris-Sud 11 and CNRS (UMR 8617), F-91405 Orsay, France}

\altaffiltext{25}{SETI Institute, 515 N. Whisman Avenue, Mountain
View, CA, 94043, USA}

\altaffiltext{26}{Physics Department, University of Johannesburg,
P.O. Box 524, Auckland Park, 2006, South Africa}

\altaffiltext{27}{Institut d'Astrophysique de Paris, CNRS and
  Universit\'{e} Pierre et Marie Curie, 98bis Bolulevard Arago,
  F-75014 Paris, France}

\altaffiltext{28}{Department of Astronomy, University of Padova,
Vicolo dell'Osservatorio 3, I-35122 Padova, Italy}

\altaffiltext{29}{Department of Physics and Astronomy, University of
British Columbia, Vancouver, BC V6T 1Z1, Canada}

\altaffiltext{30}{Astrophysics Branch, NASA Ames Research Center, Mail
Stop 245-6, Moffett Field, CA 94035, USA}

\altaffiltext{31}{Herschel Science Centre, ESAC, ESA, PO Box 78,
Villanueva de la Canada, 28691 Madrid, Spain}

\altaffiltext{32}{Leiden Observatory, Leiden University, PO Box 9513,
  NL - 2300 RA Leiden, The Netherlands}

\altaffiltext{33}{Oxford Astrophysics, Denys Wilkinson Building,
  University of Oxford, Keble Road, Oxford, OX1 3RH, UK}

\begin{abstract}

The {\em Herschel} Astrophysical Terahertz Large Area Survey (H-ATLAS)
has uncovered a population of strongly-lensed submillimeter galaxies
(SMGs).  The Zpectrometer instrument on the Green Bank Telescope (GBT)
was used to measure the redshifts and constrain the masses of the cold
molecular gas reservoirs for two candidate high-redshift lensed
sources.  We derive \coa redshifts of $z=3.042\pm0.001$ and
$z=2.625\pm0.001$, and measure molecular gas masses of (1--3)$\times
10^{10}\msun$, corrected for lens amplification and assuming a
conversion factor of $\alpha = 0.8 \msun(\Kkpspc)^{-1}$.  We find
typical $L({\rm IR})/L^{\prime}(\co)$ ratios of 120$\pm$40 and $140\pm
50 \lsun (\Kkpspc)^{-1}$, which are consistent with those found for
local ULIRGs and other high-redshift SMGs.  From analysis of published
data, we find no evidence for enhanced $L({\rm IR})/L^{\prime}(\coam)$
ratios for the SMG population in comparison to local ULIRGs.  The GBT
results highlight the power of using the CO lines to derive blind
redshifts, which is challenging for the SMGs at optical wavelengths
given their high obscuration.

\end{abstract}


\section{Introduction}

The discovery of the sub-millimeter galaxies (SMGs) thirteen years ago
revolutionized our understanding of galaxy formation and evolution by
uncovering a population of high-redshift, dust-obscured systems that
are forming stars at tremendous rates (Smail et al. 1997).  In terms
of their infrared luminosities, the SMGs are analogous to the local
ultraluminous infrared galaxies (ULIRGs).  Until recently, deep
observations at far-infrared (FIR)/sub-mm wavelengths have been either
limited to relatively small areas of the sky or severely affected by
source confusion due to poor spatial resolution.  The {\em Herschel}
Space Observatory (Pilbratt et al.\ 2010) has enormously extended the
sky coverage at FIR/sub-mm wavelengths.  The {\em Herschel}
Astrophysical Terahertz Large Area Survey (H-ATLAS, Eales et al.\
2010) will map 570 square degrees in five bands from 100 to
500$\,\mu$m.  As pointed out by Blain (1996), the sub-mm band is well
suited for generating large samples of strongly lensed galaxies at
high redshift due to the large negative $k$-correction and steep
source counts.  The Science Demonstration Phase (SDP) H-ATLAS
observations have confirmed the excess of bright lensed SMGs over the
expected number counts of unlensed galaxies (Negrello et al.\ 2010),
which is consistent with the results found from the South Pole
Telescope Survey (Vieira et al.\ 2010; Lima et al.\ 2010).

Deriving redshifts for the lensed SMGs is challenging using
traditional optical and near-infrared techniques.  The SMGs themselves
are highly obscured, and the foreground lensing galaxies dominate the
emission seen at optical and near-infrared wavelengths.  However, we
are no longer limited to these traditional techniques.  The new
generation of wide-bandwidth spectrometers operating at cm/mm/sub-mm
wavelengths now make it possible to determine redshifts directly from
the CO rotational lines (e.g., Wei\ss\, et al. 2009; Swinbank et
al. 2010).  In addition to accurate redshift measurements, CO
observations are fundamental to our understanding of galaxy evolution
by measuring the mass of the molecular gas reservoir from which stars
form.  The first two SMGs in which CO was detected (Frayer et al.\
1998, 1999) were relatively bright at optical wavelengths, which
enabled timely follow-up CO observations.  Subsequent CO detections of
additional SMGs took several years.  In general, deep radio or
mm/sub-mm wavelength interferometric continuum maps were required to
derive accurate counterpart positions (Frayer et al.\ 2000; Ivison et
al.\ 2002; Dannerbauer et al.\ 2002; Younger et al.\ 2009).  Then,
deep spectroscopic observations were necessary to obtain redshifts
(Ivison et al.\ 1998; Barger et al.\ 1999; Frayer et al.\ 2003;
Chapman et al.\ 2005).  With accurate redshifts, CO observations were
finally possible (e.g., Neri et al.\ 2003; Greve et al.\ 2005).
Hence, the process of following-up SMGs in CO used to be very time
consuming.  Now, using the wide-bandwidth radio spectrometers, we can
bypass the intermediate steps and directly search for CO lines at the
location of the SMGs uncovered by {\em Herschel} and other sub-mm
instruments.  An additional important advantage of direct CO searches
is that we avoid the biases related to the radio and optical selection
of candidate SMG counterparts.

In this paper, we report on \coa observations of two lensed SMGs
(SDP.81 and SDP.130) uncovered by the H-ATLAS program using the
wide-bandwidth Zpectrometer instrument on the Robert C. Byrd Green
Bank Telescope (GBT) operated by the National Radio Astronomy
Observatory (NRAO).  A cosmology of ${\rm H}_0 = 70\kps\,{\rm
Mpc}^{-1}$, $\Omega_{\rm M}=0.3$, and $\Omega_{\Lambda}=0.7$ is
assumed throughout this paper.

\section{Observations}

The discovery of the lensed SMGs SDP.81 (H-ATLAS J090311.6+003906) and
SDP.130 (H-ATLAS J091305.0$-$005343) was reported by Negrello et al.\
(2010).  Both of these sources have spectral-energy distributions
(SEDs) that peak in the 350$\mu$m band, suggesting redshifts of $z\sim
2.5$--3 assuming local ULIRG SEDs that peak at 80--100$\mu$m in their
rest frame.  Given their strong observed-frame 350$\mu$m emission
(180\,mJy and 130\,mJy for SDP.81 and SDP.130 respectively), they were
ideal targets for the Zpectrometer instrument on the GBT.  The
instrument currently covers the 25.61--36.09 GHz band corresponding to
redshifted \coa from $z=2.19$--3.50.  The Zpectrometer is an analog
lag cross-correlator spectrometer connected to the GBT dual-channel
Ka-band correlation receiver (see Harris et al.\ 2010 for additional
details).

The Zpectrometer observations of SDP.81 and SDP.130 were carried out
on 2010 March 24 and April 21 (GBT program 10A-77).  The observations
were taken using the sub-reflector beam switching (``SubBeamNod'')
mode with a 10 second switching interval.  Alternating sets of
SubBeamNod observations between the two targets were taken every 4
minutes to remove the residual baseline structure.  By differencing
the resulting spectra of the two targets SDP.81 (``ON'' position) and
SDP.130 (``REF'' position) flat baselines were achieved (Fig. 1).  We
obtained 2.3 hours of effective integration time (1.15 hours of ON
time and 1.15 hours of REF time).  During the observing cycle, the
observations were about 65\% efficient.  Including the additional time
for pointing, focus, and calibration, 5.5 hours of telescope time were
used.  Spectra of the pointing source 0825+0309 were taken every hour
to monitor gain variations.  The absolute flux density scale was
derived from observations of 3C286.  We adopt a flux density of 2.04
Jy at 32 GHz for 3C286 (Ott et al.\ 1994).  After correcting for
atmospheric opacity effects and based on the dispersion of
measurements observed for 0825+0309, we estimate a 15\% absolute
calibration uncertainty for the data.


\section{Results}

We observe strong emission lines in the spectra of both SDP.81 and
SDP.130 which we identify as \coa, yielding redshifts of $z=3.042$ and
$z=2.625$, respectively (Fig. 1).  Follow-up \coc observations made
with the Plateau de Bure Interferometer (PdBI) (R. Neri et al.\ 2010,
in preparation) and the higher level CO transitions observed with the
Caltech Submillimeter Observatory (CSO) Z-Spec instrument (Lupu et
al. 2010) confirm the \coa line identifications with the GBT.
Single-component Gaussian fits to the lines were made to derive the
\coa properties (Table~1).  The \coa profiles are slightly asymmetric
in directions consistent with the PdBI \coc profiles, but given the
limited \coa spectral resolution, the Zpectrometer profiles are
consistent with a single Gaussian.  The instrumental spectral response
is nearly a sinc function with a full-width half max (FWHM) of 20 MHz.
For intrinsic Gaussian FWHM line widths larger than 30 MHz ($\sim
300\kps$), the line width correction for the instrumental response is
less than 1\%.  Figure~2 shows the Gaussian fits to the raw 8 MHz
channels.  The raw channels were binned by three channels to yield
statistically independent channels.  We achieved an rms of 0.18 mJy
per 24 MHz channel, albeit with significant variation across the full
Zpectrometer bandwidth.  The observations were taken at fixed
frequencies (topocentric velocity scale).  Small Doppler corrections
of $35 \kps$ were applied to the observed line centers to derive the
redshifts with respect to the local standard of rest (LSR, Table~1).

The mass of molecular gas (including He) is computed using
$M(\hh)/L^{\prime}(\coam) = \alpha$, where the \coa line luminosity is
in units of $\Kkpspc$ (Solomon et al. 1997).  While there is still
considerable discussion about the appropriate value for $\alpha$ at
high-redshift (e.g. Tacconi et al. 2008; Ivison et al. 2010a), we
adopt the ``standard'' local ULIRG value of $\alpha=0.8
\msun(\Kkpspc)^{-1}$ (Downes \& Solomon 1998) for direct comparison
with previous studies.  However, a higher value may be more
appropriate for the expected multi-phase molecular ISM in strong
starburst systems (Papadopoulos et al.\ 2007; Harris et al.\ 2010;
Danielson et al.\ 2010; Ivison et al.\ 2010a).

\section{Discussion}

\subsection{CO Properties}

Observations of \coa are key for deriving the total molecular gas
mass.  The \coa line traces the cold material not probed by the
higher-level CO transitions.  Unfortunately, most high redshift
sources to date have been observed in only the higher level $J_{\rm
upper}>1$ transitions (e.g., \coc or higher), and many papers have
assumed a single-component ISM that is fully thermalized with
$T_{b}(J_{\rm upper}>1)/T_{b}(\otz) = L^{\prime}(J_{\rm
upper}>1)/L^{\prime}(\coam) = 1$.  However, recent \coa observations
of the SMG population show that this assumption is not correct (Harris
et al.\ 2010; Carilli et al.\ 2010; Swinbank et al.\ 2010; Ivison et
al.\ 2010a,b).  From the compilation of these previous results, an
average CO line ratio value of $r_{31} =
L^{\prime}(\cocm)/L^{\prime}(\coam) = 0.6\pm0.1$ is found for a sample
of nine SMGs.  SDP.81 and SDP.130 also have ratios of less than unity.
Based on the \coc observations made with the PdBI (R. Neri et al.\
2010, in preparation), SDP.81 and SDP.130 have $r_{31}$ ratios of 0.5
and 0.7, respectively.  These ratios are in agreement with the
previous SMG results, as well as the average value of $r_{31}\simeq
0.6$ measured for local infrared galaxies (Yao et al.\ 2003; Leech et
al. 2010).  Although the values measured for SMGs are similar to the
average value found for the local starburst population, there are
significant variations in the ratio locally ($r_{31} =0.1$--1).  The
possible wide range of CO line ratios highlights the importance of
obtaining \coa observations for the SMGs and other high-redshift
populations for comparison.  For example, the BzK galaxies may show
similar ``sub-thermal'' CO line ratios as the SMGs (Dannerbauer et
al. 2009; Aravena et al.\ 2010), while in contrast high-redshift
quasars tend to show CO lines ratios of order unity up to CO(4-3) or
even higher transitions (e.g., Riechers et al. 2006).

Based on their $L^{\prime}(\coam)$ luminosities corrected for
amplification by lensing (Table~1), the derived molecular gas masses
for SDP.81 and SDP.130 are (1--3)$\times10^{10}\msun$, which is about
3--5 times larger than that found for local ULIRGs (Downes \& Solomon
1998), but is consistent with other SMGs studied to date (adopting the
same $\alpha =0.8 \msun(\Kkpspc)^{-1}$).  Since the infrared
luminosity is proportional to the star-formation rate and the CO
luminosity is proportional to the molecular gas mass, the infrared to
CO luminosity ratio provides an indication of the star-formation
efficiency.  However, given the uncertainties of converting the
observables into physical quantities (especially the uncertainty of
$\alpha$), we restrict the discussion to the observed $L({\rm
IR})/L^{\prime}(\coam)$ ratios.  Strong starbursts and ULIRGs tend to
show high IR-to-CO luminosity ratios of $\ga 100 \lsun
(\Kkpspc)^{-1}$, while local spiral galaxies have lower values of
about 10--50 (e.g., Solomon \& Vanden Bout 2005).  Based on their
infrared luminosities (Table~1), SDP.81 and SDP.130 have $L({\rm
IR})/L^{\prime}(\coam)$ ratios of 120 and 140 $\lsun (\Kkpspc)^{-1}$,
respectively.

Greve et al.\ (2005) found an average $L({\rm IR})/L^{\prime}(\co)$
ratio for SMGs which was a factor of two larger than that for local
ULIRGs.  However, if the published data are corrected for the same
cosmology (Sec. 1), infrared luminosity definition [$L({\rm IR})$,
8--1000$\mu$m], and for the same \coa transition, the SMGs actually
show a slightly lower $L({\rm IR})/L^{\prime}(\coam)$ ratio on average
in comparison to local ULIRGs.  We recomputed the infrared
luminosities for the local ULIRG sample given by Solomon et al.\
(1997), using the relationship in Sanders \& Mirabel (1996), and find
a median value of $L({\rm IR})/L^{\prime}(\coam) = (240\pm 60) \lsun
(\Kkpspc)^{-1}$ (for $L({\rm IR}) > 10^{12} \lsun$).  The uncertainty
represents the standard deviation of the ratio for the sample after
throwing out one outlier, divided by the square-root of the number of
sources (28), and includes an additional 15\% systematic calibration
uncertainty.  Adopting $r_{31} =0.6$, the implied $L^{\prime}(\coam)$
values for SMGs are increased by a factor of 1.67 in comparison to
Greve et al.\ (2005) which assumed $r_{31} =1$.  The average $L({\rm
IR})$ value from Greve et al.\ (2005) also needs to be decreased by
roughly a factor of 2 due to the lower observed dust temperatures
measured for the SMGs (Kov\'{a}cs et al.\ 2006; Magnelli et al. 2010).
After making these corrections, the Greve et al.\ (2005) results
suggest an average value of $L({\rm IR})/L^{\prime}(\coam) = 110 \pm
40 \lsun (\Kkpspc)^{-1}$ for the SMG population.  This value is
consistent with those found for SDP.81 and SDP.130, and suggests that
SMGs do not have enhanced $L({\rm IR})/L^{\prime}(\coam)$ ratios in
comparison to local ULIRGs.

We carried out an additional comparison with the local ULIRGs, by
using only SMGs with infrared measurements near the peak of their
rest-frame SED (60--180$\mu$m) (Fig. 3).  This improves the
derivations of $L({\rm IR})$ that are dependent on the assumed SED
template.  Including SDP.81 and SDP.130 with a sample of published
SMGs (Fig. 3), we derive a median value of $L({\rm
IR})/L^{\prime}(\coam) = 125 \pm 50 \lsun (\Kkpspc)^{-1}$ for the SMG
population.  For sources without \coa detections, the published data
were converted to \coa luminosities adopting
$r_{21}=L^{\prime}[\cobm]/L^{\prime}[\coam] =0.8\pm0.1$,
$r_{31}=L^{\prime}[\cocm]/L^{\prime}[\coam] =0.6\pm0.1$, and
$r_{41}=L^{\prime}[\codm]/L^{\prime}[\coam] =0.4\pm0.1$.  The adopted
$r_{41}$ and $r_{21}$ values are consistent with simple large-velocity
gradient excitation analysis based on $r_{31}=0.6$ (with temperatures
of 20--50\,K and molecular gas densities of order 1000\,cm$^{-3}$) and
are consistent with current observations of local luminous starbursts.

A direct comparison of the $L({\rm IR})/L^{\prime}(\coam)$ ratios with
that of local ULIRGs is complicated by the possibility of strong
mid-infrared emission (rest frame 25$\mu$m).  The SMG templates used
to derive $L({\rm IR})$ do not include possible excess mid-infrared
emission, which is present in some local ULIRGs.  However, if we
neglect the mid-infrared emission for local ULIRGs, the median $L({\rm
IR})/L^{\prime}(\coam)$ ratio would decrease by only 10\%.  Including
this small possible correction for mid-infrared emission, the median
$L({\rm IR})/L^{\prime}(\coam)$ value for SMGs is $0.6\pm0.3$ times
that found for local ULIRGs, i.e., the median ratio is slightly lower
in SMGs, but roughly consistent with local ULIRGs within
uncertainties.

The observed $r_{31}$ brightness temperature ratios and the $L({\rm
IR})/L^{\prime}(\co)$ luminosity ratios for SDP.81 and SDP.130 are
similar to values found for ULIRGs and other SMGs.  Since the \coa
emission traces the cold gas, which is typically more spatially
extended than the \coc emission (Ivison et al.\ 2010a), differential
lensing could impact the interpretation of the results for SDP.81 and
SDP.130.  If differential lensing is important, then the intrinsic
$r_{31}$ ratios for SDP.81 and SDP.130 may be even lower than the
values derived here.

\subsection{CO Redshift Surveys}

The new generation of broad bandwidth spectrometers now enables blind
redshift searches for CO emission (see conference proceedings of Baker
et al. 2007).  The first blind redshift for the Zpectrometer
instrument was found for SMM J2135-0102 (Swinbank et al.\ 2010).
SDP.81 and SDP.130 represent additional blind redshifts from the
Zpectrometer.  The first blind CO redshift for the Caltech
Submillimeter Observatory (CSO) Z-Spec instrument was found for SDP.81
(Lupu et al.\ 2010).  Wei\ss\, et al.\ (2009) reported the first blind
CO redshift using the Institut de Radioastronomie Millim\'{e}trique
(IRAM) 30m Eight MIxer Receiver (EMIR) instrument, and Daddi et
al. (2009) and M. Krips et al.\ (2010, in preparation) have
successfully used the PdBI to uncover previously unknown redshifts
with CO lines.  Table~2 shows the capabilities of the current and
planned instrumentation for CO redshift machines.  Based on the
combination of its large fractional bandwidth (34\%) and good
sensitivity, the GBT/Zpectrometer is currently the most efficient
system for searching for \coa lines at redshifts $2.2<z<3.5$.
Although not as sensitive as the GBT, the Z-Spec instrument has a
larger fractional bandwidth (46\%) and can search all redshifts using
a variety of transitions.  At the highest frequencies, CO searches may
be difficult due to subthermal excitation of the high-J CO lines, and
the atomic lines such as [CI] and [CII] may be more feasible (e.g.,
Wagg et al. 2010).  A primary science driver for the Large Millimeter
Telescope (LMT) is CO redshift searches at 3\,mm using the Redshift
Search Receiver (RSR, Erickson et al.\ 2007).  When the Expanded Very
Large Array (EVLA) achieves its full 8~GHz of bandwidth, it will be
able to search for \coa at redshifts $z>1.3$ with better sensitivity
than that of the GBT/Zpectrometer (Table 2).

\section{Concluding Remarks}

The sub-millimeter galaxies SDP.81 and SDP.130 are two of the first
examples of the lensed SMG population discovered by the {\em Herschel}
Space Observatory (Negrello et al.\ 2010).  They have CO properties
similar to those found for other high-redshift SMGs and local ULIRGs
in terms of their CO line ratios and their infrared to CO luminosity
ratios.  In contrast to previous results, we find no evidence for
enhanced $L({\rm IR})/L^{\prime}(\co)$ ratios for the SMGs in
comparison to local ULIRGs.  Given their high amplification, the {\em
Herschel} population of lensed SMGs provides ideal targets for
studying the ISM properties at high redshift, by allowing observations
of fainter lines, such as HCN, $^{13}$CO, and [CI], which would
otherwise be too faint.  Studying multiple molecular species and
detailed imaging of several CO transitions are required to constrain
the different components of the molecular ISM at high redshift and
their CO to H$_2$ conversion factors.

In the upcoming era of high-resolution imaging with the Atacama Large
Millimeter/submillimeter Array (ALMA) and the EVLA, large single
dishes will still have a major role to play in spectroscopic CO
surveys.  The GBT, LMT, and eventually the Cerro Chajnantor Atacama
Telescope (CCAT) will be able to determine redshifts for significant
samples of highly obscured SMGs which are not measureable with even
the largest optical and near-infrared telescopes.

\acknowledgments

We acknowledge the staff at Green Bank who have made these
observations possible.  We are indebted to the late Senator Robert
C. Byrd for his strong support of the Green Bank Telescope.  The
National Radio Astronomy Observatory is a facility of the National
Science Foundation operated under cooperative agreement by Associated
Universities, Inc.  AJB acknowledges support from the National Science
Foundation through grant AST-0708653.

{\it Facility:} \facility{GBT (Zpectrometer)}

%

\begin{table}
\begin{center}
\caption{\coa Observational Results}
\tablewidth{300pt}
\begin{tabular}{lcc}
\tableline\tableline
Parameter & SDP.81 & SDP.130\\
\tableline

Line Peak $S_{\nu}$ [mJy] & 2.39$\pm$0.19  & 1.63$\pm$0.22 \\

FWHM $\Delta V$ [$\kps$] & 435$\pm$54 & 377$\pm$62\\

Integrated Line Flux $S({\rm CO})$ [$\jy \kps$]\tablenotemark{a} &
1.11$\pm$0.25 &0.65$\pm$0.19\\
  
Line Center (topocentric) [GHz] & 28.515$\pm$0.003 & 31.798$\pm$0.005\\

Redshift [z(LSR)] & 3.042$\pm$0.001 & 2.625$\pm$0.001\\

$L(\co)$ [$10^{6} \lsun$]\tablenotemark{b} & $0.9\pm0.3$  & $1.7\pm0.5$ \\

$L^{\prime}(\co)$ [$10^{10} \Kkpspc$]\tablenotemark{b} & $1.8\pm0.7$ &
$3.4\pm1.0 $\\

$M(\hh)$ [$\msun$]\tablenotemark{b,c} & $\sim 1.4\times 10^{10}$ & $\sim
2.7\times 10^{10}$ \\
 
$L({\rm IR})$ [$ 10^{12}\lsun$]\tablenotemark{b,d}& $2.1\pm 0.7$ &
$4.7\pm1.3$ \\

$L({\rm IR})/L^{\prime}(\coam)$ [$\lsun (\Kkpspc)^{-1}$]\tablenotemark{e} & 120$\pm$40 & 140$\pm$50 \\

\tableline
\end{tabular}

\tablenotetext{a}{Uncertainty on the total \coa line flux includes the
  15\% systematic calibration uncertainty added to the statistical
  noise of the line.}  \tablenotetext{b}{Corrected for the lensing
  amplification factors of $25\pm7$ for SDP.81 and $6\pm1$ for SDP.130
  (Negrello et al.\ 2010).}  \tablenotetext{c}{Adopting $\alpha = 0.8
  \msun(\Kkpspc)^{-1}$ (Downes \& Solomon 1998) which could be
  uncertain by a factor of 2 or more.}
  \tablenotetext{d}{$L(IR)$[8--1000$\mu$m] based on fitting an Arp 220
  template.}  \tablenotetext{e}{Assuming no differential lensing
  between the CO and infrared emission.}

\end{center}
\end{table}

\begin{table}
\begin{center}
\caption{Instruments for CO Redshift Searches}
\tablewidth{300pt}
\begin{tabular}{lcccc}
\tableline\tableline

Telescope & Instrument & Frequency Range & Bandwidth & Sensitivity
($5\sigma$)\tablenotemark{a}\\

\tableline
GBT  & Zpectrometer & 25.6 -- 36.1 GHz  & 34\% &0.9 mJy (this work)\\
CSO  & Z-Spec & 190 -- 305 GHz & 46\% & 100 mJy (Lupu et al. 2010)\\
CSO  & ZEUS\tablenotemark{b}   & 632 -- 710 GHz & 4\% & 300 mJy
(Ferkinhoff et al. 2010) \\
IRAM 30m & EMIR\tablenotemark{b} & 83 -- 117 GHz   & 8\% & 9 mJy
(IRAM documentation)\\
PdBI & WideX\tablenotemark{b}  &  80 -- 116 GHz   & 3.6\% & 3.7 mJy
(Daddi et al. 2009)\\
CARMA\tablenotemark{b,c} &     & 85 -- 116 GHz    & 8\%& 13 mJy (web calculator)\\

EVLA\tablenotemark{c} & WIDAR  & 12 -- 50 GHz     &
40--18\% & 0.2--0.4 mJy (project page) \\

LMT\tablenotemark{d}  &RSR &74 -- 111 GHz  & 40\% & 4
mJy (32m), 1.5 mJy (50m)\tablenotemark{e}\\     
ALMA\tablenotemark{b,d} &         & 84 -- 116 GHz & 8\% & 0.4 mJy (web
calculator) \\
\tableline
\end{tabular}
\tablenotetext{a}{Estimated $5\sigma$ sensititivity with 1 hour of on
source integration for a line width of $300\kps$.  Please consult
observatory documentation for updated estimates of sensitivity.}
\tablenotetext{b}{Higher frequency bands with lower fractional
bandwidth also available.}  \tablenotetext{c}{The Combined Array for
Research in Millimeter Astronomy (CARMA) has a mixture of 4 GHz and 8
GHz bandwidths.}  \tablenotetext{d}{System still in development.}
\tablenotetext{e}{The LMT sensitivity estimated for the initial 32m
  telescope and the final 50m telescope.}

\end{center}
\end{table}

\begin{figure}
\epsscale{0.9}
\plotone{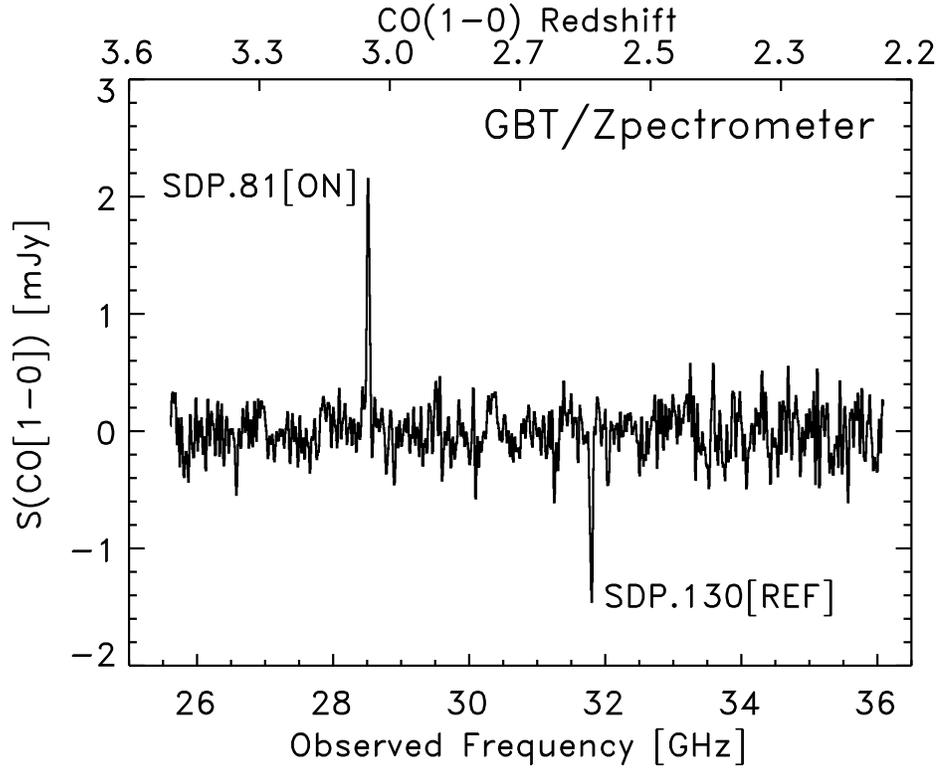}
\vspace*{1cm}
\caption{The full GBT/Zpectrometer difference spectrum for SDP.81 (ON
  source) and SDP.130 (REF source) showing the detections of \coa
  emission for both. SDP.81 was detected at the 12$\sigma$ level, while
  SDP.130 was detected at 7$\sigma$ (Table~1).}
\end{figure}

\begin{figure}
\epsscale{1.0}
\plottwo{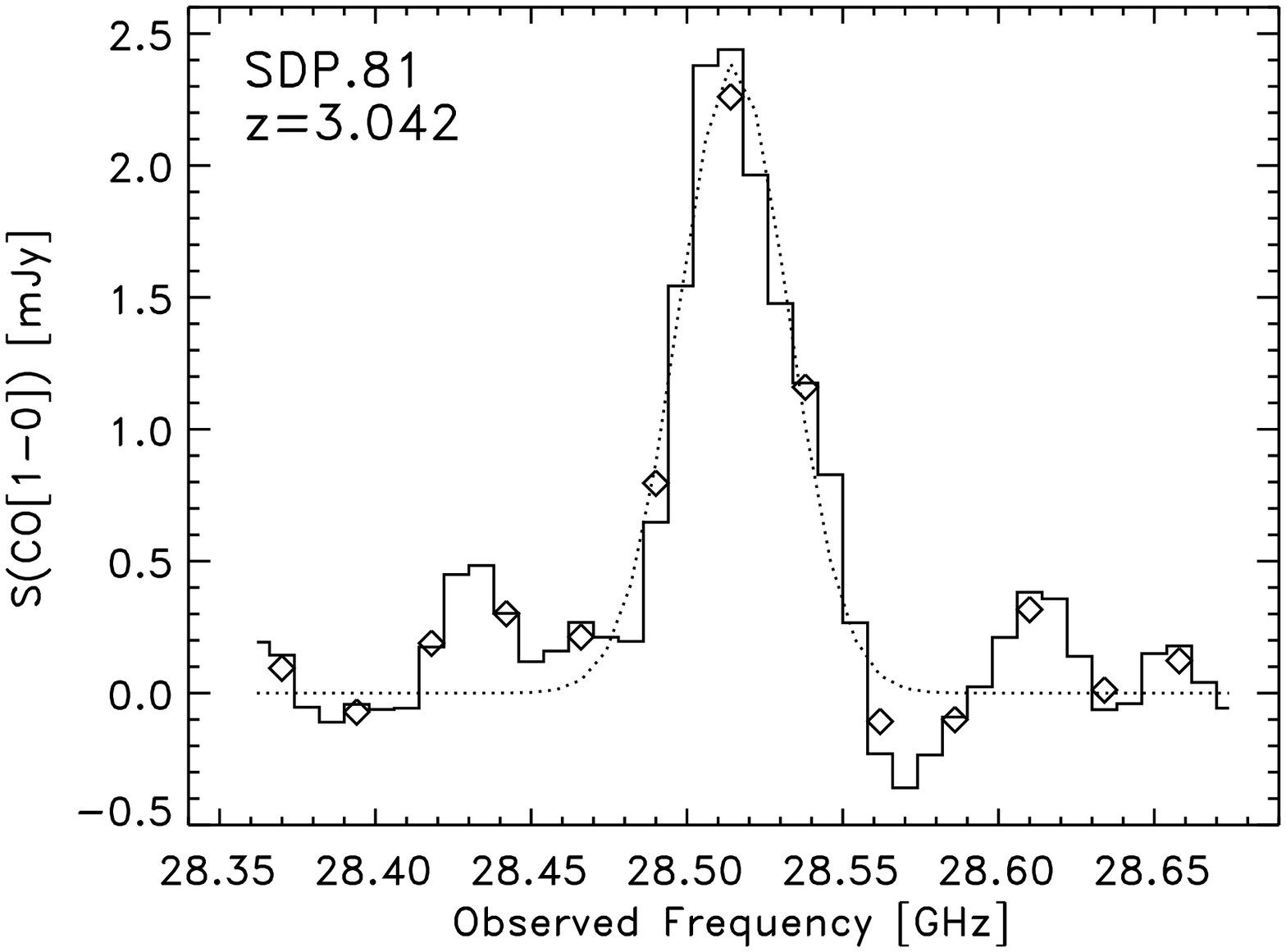}{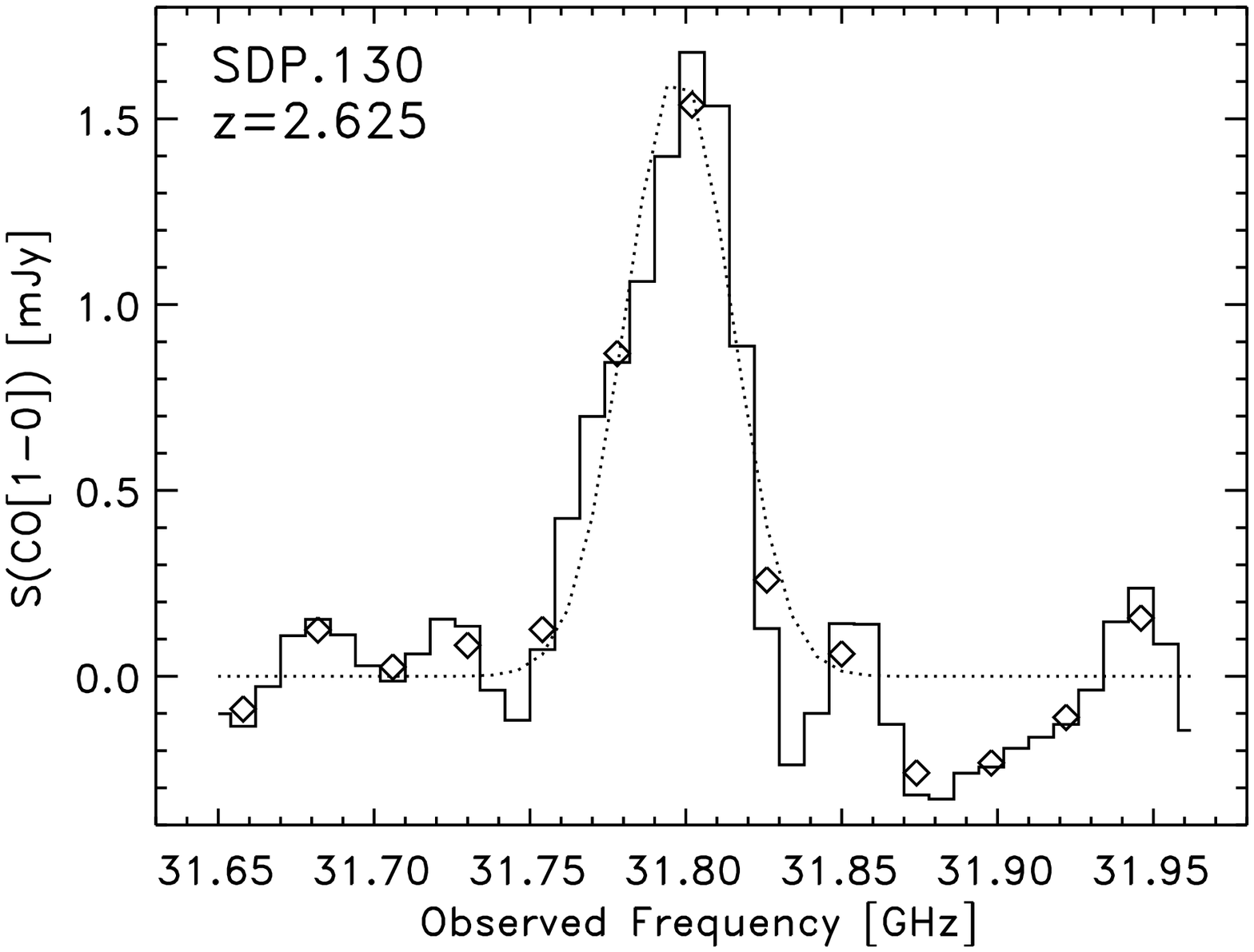}
\vspace*{1cm}
\caption{The \coa spectra for SDP.81 (left) and SDP.130 (right).  The
solid-line histograms show the raw data, and the Gaussian fits to the
raw data are shown by the dotted lines.  The diamonds represent
independent 24 MHz channels derived by binning the raw data by 3
channels.}
\end{figure}
\begin{figure}

\epsscale{0.8}
\plotone{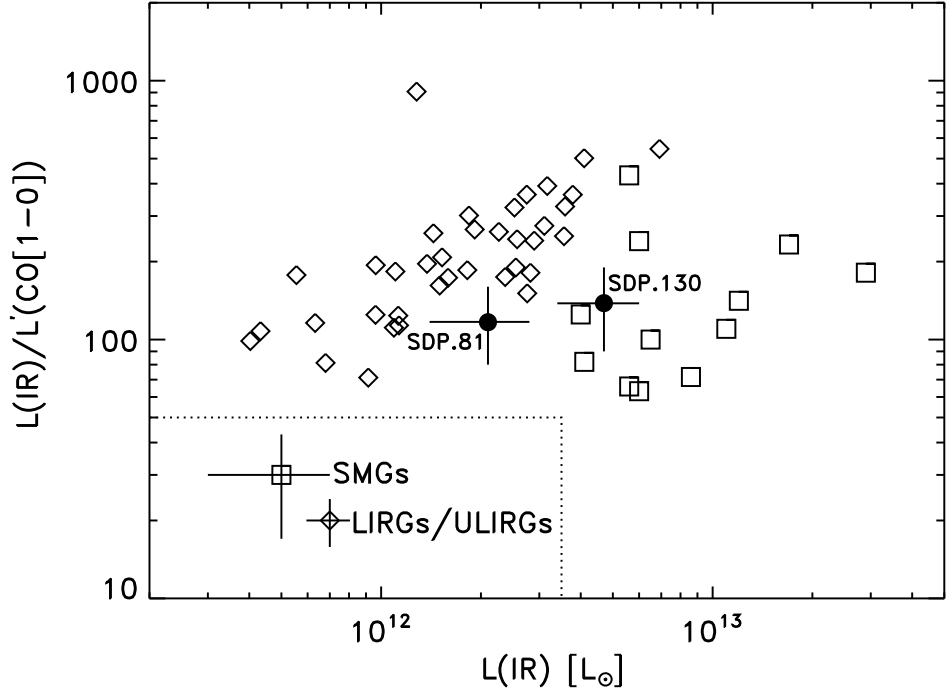}
\vspace*{1cm}
\caption{The $L({\rm IR})/L^{\prime}(\coam)$ luminosity ratio in units
  of $\lsun (\Kkpspc)^{-1}$ for local LIRGs/ULIRGs given by Solomon et
  al.\ (1997) [diamonds] and a sample of SMGs [squares] (Greve et al.\
  2003, 2005; Kov{\'a}cs et al.\ 2006; Solomon \& Vanden Bout 2005;
  Frayer et al.\ 2008; Daddi et al.\ 2009; Carilli et al.\ 2010;
  Ivison et al.\ 2010b; Coppin et al.\ 2010; Riechers et al.\ 2010).
  All SMGs were chosen based on the existence of infrared measurements
  near the peak of their rest-frame SED (60--180$\mu$m).  Based on the
  CO line ratios found for the SMGs and ULIRGs, the published data
  were converted to \coa luminosities adopting $r_{21}=0.8$,
  $r_{31}=0.6$, and $r_{41}=0.4$ for the SMGs without \coa detections.
  All points assume the same cosmology and infrared luminosity
  definition (8--1000$\mu$m).  The SMGs SDP.81 and SDP.130 are shown
  by the solid points, and the approximate errors for the other points
  are given by the crosses at the lower left.}
\end{figure}

\end{document}